\documentclass[a4paper,12pt,oneside,fleqn]{article}
\usepackage{amsmath,amssymb,amsfonts}
\usepackage{graphicx}
\usepackage{psfrag}

\usepackage[margin=1.0in]{geometry}
\usepackage[font=small,margin=0.5in]{caption}
\usepackage{setspace}

\newcommand{\df}[1]{\delta\!\left({#1}\right)}

\newcommand{\upd}{\mathrm{d}}

\begin{document}

\doublespacing

\noindent {\bf \Large Condensation Transition in Fat-Tailed Distributions:}\\
\noindent {\bf \Large a Characterization by Means of an Order Parameter}\\

\onehalfspacing

\indent {\bf Mario Filiasi}\\
\indent Physics Department, University of Trieste,\\ \indent via Valerio 2, I-34127 Trieste, Italy;\\
\indent LIST S.p.A., via Carducci 20, I-34122 Trieste, Italy.\\

\indent {\bf Elia Zarinelli}\\
\indent LIST S.p.A., via Carducci 20, I-34122 Trieste, Italy.\\

\indent {\bf Erik Vesselli\footnote{Corresponding author: vesselli@iom.cnr.it}}\\
\indent Physics Department and CENMAT, University of Trieste,\\ \indent via Valerio 2, I-34127 Trieste, Italy;\\
\indent IOM-CNR, Laboratorio TASC, Area Science Park,\\ \indent S.S. 14 km 163.5, I-34149 Basovizza (Trieste), Italy.\\

\indent {\bf Matteo Marsili}\\
\indent The Abdus Salam International Centre for Theoretical Physics (ICTP),\\ \indent Strada Costiera 11, I-34014 Trieste, Italy.\\

Condensation phenomena are ubiquitous in nature and are found in condensed matter, disordered systems, networks, finance, etc. In the present work we investigate one of the best frameworks in which condensation phenomena take place, namely, the sum of independent and fat-tailed distributed random variables. For large deviations of the sum, this system undergoes a phase transition and shifts from a \emph{democratic} phase to a \emph{condensed} phase, where a single variable (the condensate) carries a finite fraction of the sum. This phenomenon yields the failure of the standard results of the Large Deviation Theory. In this work we exploit the Density Functional Method to overcome the limitation of the Large Deviation Theory and characterize the condensation transition in terms of an order parameter, i.e.\ the Inverse Participation Ratio (IPR). This procedure leads us to investigate the system in the large-deviation regime where both the sum and the IPR are constrained, observing new phase transitions. As a sample application, the case of condensation phenomena in financial time-series is briefly discussed.

\section{Introduction}
\label{sec:intro}

The ubiquity of normal-distributed observables in natural sciences can be understood, in some sense, in terms of the Central Limit Theorem (CLT). The CLT states that the distribution of the sum of independent and identically distributed (i.i.d.)\ random variables converges, for a large number of variables, to a normal distribution. This statement does not depend on the details of the variables' distribution but only on the existence of their variance, and this explains why normal distributions are so widespread in nature. The CLT has been widely generalised, allowing to identify many other universal limiting distributions. The sum of i.i.d.\ random variables with infinite variance is shown to converge to a L\'evy stable distribution. At variance, the maximum of i.i.d.\ random variables converges to a Gumbel, a Fr\'echet, or a Weibull distribution, depending on the tails of the vairiables' distribution. An interesting phenomenon concerning both the sum and the maximum of i.i.d.\ random variables takes place in the uncommon region of very broad distributions with indefinite mean (those distibutions which violate the Law of Large Numbers). In this region, the statistical properties of the sum and of the maximum share common features: their fluctuations are so strong that, in some sense, a finite fraction of the sum is carried by a single variable, the maximum. This phenomenon is usually referred to as \emph{condensation}.

Interestingly, condensation is not exclusive of very broad distributions with indefinite mean, but is a much more general phenomenon. It has been recently shown \cite{majumdar2005nature,marsili2012concentration} that, if we investigate the proper region of the phase space, we can observe condensation for any \emph{fat-tailed} distribution (i.e.\ any distribution whose tails decay slower than an exponential), even if its moments are finite. Specifically, condensation phenomena appear in the \emph{large deviations} regime of the sum of all fat-tailed random variables, and characterize their untypical outcomes. This phenomenon has profound implications on the Large Deviations Theory (LDT), which aims at extimating the probability of rare events \cite{touchette2009large}. For fat-tailed random variables the main results of the LDT break down (the rate function vanishes, the rare events are not exponentially suppressed, etc.), and the theory gives no predictions about how the untypical outcomes are realized and what their probability is. In the present work we exploit a formalism borrowed from the Random Matrix Theory, which allows us to overcome these limitations. It allows us to recover all common results of the LDT, as well as to obtain a new detailed description of condensation phenomena in the regime where the standard LDT stops working. Our results are in agreement with the ones reported in \cite{majumdar2005nature}, where condensation phenomena are studied in the context of mass transport models.

Condensation phenomena in fat-tailed random variables manifest through a phase transition between a \emph{democratic} and a \emph{condensed} phase \cite{majumdar2005nature,marsili2012concentration}. Here, with a statistical-mechanics flavour, we investigate the order parameter for the transition, the so-called Inverse Participation Ratio, borrowed from solid-state physics. This leads us to investigate new large-deviations regimes where additional constraints are imposed to the system. Our results, in agreement with the recent study reported in \cite{szavits-nossan2014constraint}, suggest that condensation phenomena are very general and can be induced (or inhibited) by imposing the proper scaling-laws to system's observables.

Our object of study is of general interest.
Fat-tailed distributions, indeed, have been found to describe the probability of events in many domains: the magnitude of earthquakes \cite{zaliapin2005approximating, saichev2006universal},  forest-fires \cite{malamud1998forest}, rain events \cite{peters2002rain}, cities' size \cite{zipf1949human}, economic wealth \cite{pareto1964cours} and price returns of stocks' indices \cite{mantegna1995scaling,embrechts1997modelling} among the others. Due to the nature of fat-tailed distributions, the extreme events in such domains are not so rare and usually occur through condensed outcomes (financial crashes, hurricanes, billionaires, etc.). Moreover, condensation phenomena are not exclusive of i.i.d.\ random variables, but have been observed in a large variety of physical systems. The Bose-Einstein condensation \cite{pitaevskii2003bose} is probably the best known example in condensed-matter physics. In disordered systems, the Random Energy Model displays a condensation transition between a paramagnetic and a spin-glass phase \cite{mezard2009information}. Financial correlation matrices \cite{bouchaud2003theory}, bipartite quantum systems \cite{nadal2010phase}, networks \cite{bianconi2001bose}, and non-equilibrium mass transport models \cite{majumdar2005nature} are examples of systems in which condensation phenomena can be observed. Nevetheless, because of the lack of interactions, i.i.d.\ random variables are probably the simplest system in which condensation takes place, so they are the best framework to investigate condensation phenomena in a statistical-mechanics approach.

The application of our study to natural and social sciences are straightforward.
Let us consider, for example, the rain events on a specific city. If these events were i.i.d.\ random variables, whenever we observe an extremely large rainy year we should expect, according to our results, that this is due to a single day with exceptionally heavy rain, rather than to a large number of commonly rainy days. This is the condensation phenomenon in practice. In the present work we compare our results with empirical observations taken from finance:  we consider the price-returns of some stocks in the Italian Market. We observe how time dependence generates deviations from the expected behaviour predicted by our analysis. We observe also that, removing time correlation, our predictions are recovered.

The present work is structured as follows.
In Sec.\ \ref{sec:setting} we review some common results about the condensation transition for power-law distributions; we define condensation phenomena and we introduce the Inverse Participation Ratio as an order parameter for the phase transition.
In Sec.\ \ref{sec:largedev} we investigate the large-deviations regime of the sum of i.i.d.\ random variables by means of the Density Functional Method (DFM), we derive the standard results of LDT and analyze the behaviour of the system in the condensed phase where the LDT is not predictive.
In Sec.\ \ref{sec:prices} we compare our analytical results with empirical observations on financial time series.
In Sec.\ \ref{sec:general} we extend the analysis of Sec.\ \ref{sec:largedev} to a more general case by introducing a new large-deviation constraint.
Conclusions are drawn in Sec.\ \ref{sec:conclusions}.
Please notice that the topics of this paper are addressed also in a more recent work \cite{marsili_inpr}, where the interested reader can find a further discussion about condensation phenomena, with application to finance, inference, and random matrix theory.

\section{Setting the Stage}
\label{sec:setting}

Let us consider a set $\{x_1, x_2, \dots, x_N\}$ of non-negative i.i.d.\ random variables distributed according to a power-law probability density function (p.d.f.):
\begin{equation}
p(x) \simeq \frac{A}{x^{\alpha+1}} \ ,
\label{eq:pdf}
\end{equation}
with $\alpha>0$. The Central Limit Theorem (CLT) states that the limit p.d.f.\ of the sum of i.i.d.\ random variables:
\begin{equation}
S_N = \sum_{i=1}^N x_i \ ,
\end{equation}
converges to the Gaussian distribution if $\alpha > 2$ or to a L\'evy stable distribution if $0< \alpha \leq 2$ \cite{gnedenko1968limit}. Let us consider random variables distributed according to a p.d.f. in the L\'evy basin of attraction. As reported by Bouchaud and Georges \cite{bouchaud1990anomalous}:
\begin{itemize}
\item for $0 < \alpha \leq 1$, both $\langle x \rangle$ and $\langle S_N \rangle$ are infinite and $S_N$ scales as $N^{1/\alpha}$ (or as $N \ln N$ for $\alpha =1$);
\item for $1 < \alpha \leq 2$, both $\langle x \rangle$ and $\langle S_N \rangle$ are finite, whereas $\langle x^2 \rangle$ and the variance $\langle S_N^2\rangle - \langle S_N\rangle^2$ are infinite. The difference $S_N-\langle S_N \rangle$ scales as $N^{1/\alpha}$ (or as $\sqrt{N \ln N}$ for $\alpha =2$).
\end{itemize}
It is also possible to demonstrate that the largest variable $x_\mathrm{max}$ among all variables $\{x_1, x_2, \dots, x_N \} $ for large $N$ scales as $N^{1/\alpha}$. Since both $S_N$ and $x_\mathrm{max}$ scale in the same way for i.i.d.\ random variables with very broad distributions, the typical outcome of $S_N$ may yield \emph{condensation}, i.e. it may be dominated by the single variable $x_\mathrm{max}$.

\begin{figure}
\centering
\psfrag{X}[c][c]{$\alpha$}
\psfrag{Y}[r][c]{$\langle Y_2 \rangle$}
\includegraphics[width=0.6\columnwidth]{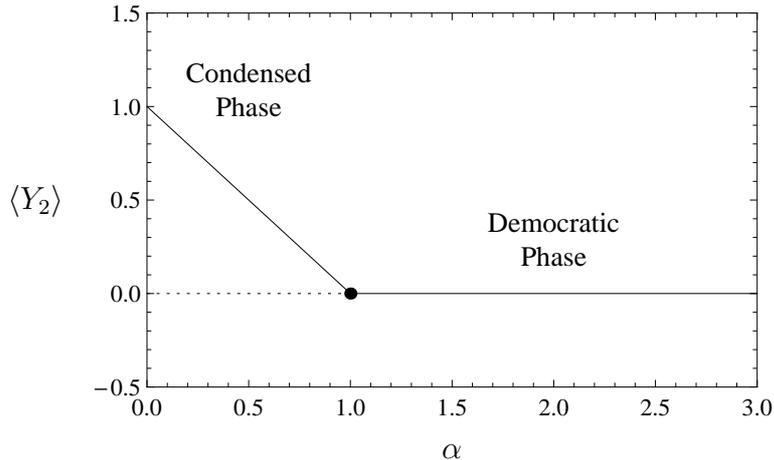}
\caption{\onehalfspacing Average value of the IPR as a function of the tail-index $\alpha$ for power-law distributions in the limit $N\to\infty$. The non-analyticity of $\langle Y_2\rangle$ at the critical point $\alpha=1$ is shown. The condensed phase is identified by non-vanishing values of the IPR.}
\label{fig_phdiag1}
\end{figure}

In order to make the previous statements more rigorous, following \cite{derrida1994non}, we can define the weight of the $i$-th term of the sum $S_N$:
\begin{equation}
w_i = \frac{x_i}{S_N} \ ,
\end{equation} 
and the $k$-th (non-centered) sample moment of the weights:
\begin{equation}
Y_k = \sum_{i=1}^N w_i^k \ ,
\end{equation}
where $k > 1$. The variable $Y_k$ can be used to quantify the degree of condensation of $S_N$, and therefore is a good candidate as an order parameter. Let us consider, for example, the second moment $Y_2$, which is called Inverse Participation Ratio (IPR). If all the $w_i$ were of order $1/N$ then $Y_2 \sim 1/N$ and would tend to zero for large $N$. On the other hand, if at least one $w_i$ remained finite when $N \to \infty$, then $Y_2$ would also be finite. We will refer to the former as a \emph{democratic outcome} and to the latter as a \emph{condensed outcome}. In this case, the variables carrying a finite weight $w_i$ of $S_N$ are called \emph{condensates}.
The average value $\langle Y_k\rangle$ can be analytically evaluated in the limit $N\to\infty$ \cite{derrida1994non}. For $\alpha \geq 1$ it is always zero, whereas for $\alpha < 1$ it reads:
\begin{equation}
\langle Y_k \rangle \simeq \frac{\Gamma(k-\alpha)}{\Gamma(k)\Gamma(1-\alpha)} \ .
\end{equation}
For $k=2$ we get $\langle Y_2 \rangle = \max\{1-\alpha,0\}$ (see Fig.\ \ref{fig_phdiag1}). These statements allows us to study the problem in a statistical mechanics flavor. The limit $N\to\infty$ plays the same role of the thermodynamic limit in physical systems. The non-analitical behavior of $\langle Y_k\rangle$ in $\alpha=1$, vanishing in the whole region $\alpha>1$, suggests that $Y_k$ is the order parameter of a phase transition. Therefore, the value $\alpha_c =1$ is a critical point for the control parameter $\alpha$ and defines two phases: a \emph{democratic phase} ($\alpha > \alpha_c$) and a \emph{condensed phase} ($\alpha < \alpha_c$). The apperance of a condensed phase is closely connected to the anomalous scaling of $S_N$ for small values of $\alpha$. Indeed, $S_N =O(N^{1/\alpha})$ for $\alpha<1$ and as $S_N =O(N)$ for $\alpha>1$.

\section{Large Deviations}
\label{sec:largedev}

While in the previous section we have shown that, for extremely fat-tailed distributions ($\alpha <1$), the \emph{typical outcomes} of $S_N$ are condensed, in the following we investigate the appearance of condensation phenomena in the \emph{large-deviations} regime of fat-tailed distributions where $S_N$ is not typically condensed ($\alpha >1$). It is well known that the CLT, when applicable, provides a good approximation only for the center of the p.d.f. of $S_N$, leaving its tails subject to further investigation. In this scenario, we can invoke the Large Deviations Theory (LDT) as an extension or refinement of the Law of Large Numbers and of the CLT \cite{touchette2009large}.

Without loss of generality, instead of the sample sum $S_N$, let us consider the sample mean $M_N = S_N/N$ which is an intensive quantity (at least for $\alpha>1$). In order to analyze the large-deviations regime of $M_N$, we consider the p.d.f.\ of the variables $\{x_1, x_2, \dots, x_N \}$ conditioned to the constraint $M_N=m$, where $m$ can be very different from its tipical value $\langle x\rangle$ (if any). This reads:
\begin{equation}
P_N(x_1,x_2,\dots,x_N|m) = \frac{\big(\prod_i p(x_i)\big)\,\df{\frac{1}{N}\sum_i x_i-m}}{P_N(M_N=m)} \ .
\label{eq:joint}
\end{equation}
The normalization constant $P_N(M_N = m)$ is the p.d.f.\ of the random variable $M_N$, denoting the probability that $M_N$ attains a value in the infinitesimal interval $[m,m+dm]$, and reads:
\begin{equation}
P_N(M_N=m) = \int \left(\prod_{i=1}^N \upd x_i\,p(x_i) \right) \df{\frac{1}{N}\sum_{i=1}^N x_i - m} \ .
\label{eq:partfunc}
\end{equation}
These equations have a specific statistical-mechanics interpretation. The distribution (\ref{eq:joint}) can be considered as a Boltzmann weight $e^{-\beta H}$ with inverse temperature $\beta=1$ and Hamiltonian:
\begin{equation}
H(x_1,x_2,\dots,x_N)=-\sum_{i=1}^N \ln p(x_i) \ ,
\label{eq:energy}
\end{equation}
under the constraint:
\begin{equation}
\frac{1}{N}\sum_{i=1}^N x_i = m \ .
\label{eq:constraint}
\end{equation}
The distribution (\ref{eq:partfunc}), at variance, is the partition function of the system depending on the external parameter $m$. Therefore, we are dealing with a system of $N$ particles subject to the potential $V(x)=-\ln p(x)$ and interacting through the global constraint (\ref{eq:constraint}). The key ingredients of the LDT is the large deviations principle, which, roughly speaking, relies on the asymptotic relation:
\begin{equation}
P_N(M_N=m) \sim e^{-NI(m)} \ , \qquad \text{for $N\to\infty$.}
\label{eq:LDT}
\end{equation}
The function $I(m)$ is the so-called \emph{rate function} and plays the same role of the free-energy density in physical systems \cite{touchette2009large}. It has been recently shown in \cite{marsili2012concentration} that, for non-negative random variables with fat-tailed distribution and finite expectation value $\langle x \rangle$, the LDT leads to a well-defined rate function only for $m<\langle x \rangle$, whereas for $m> \langle x \rangle$ the rate function vanishes and the large-deviations principle does not hold anymore. Such phenomenon implies the presence of a new phase transition which takes place in the large-deviations regime of $M_N$. The new control parameter is the mean $m$, and its critical value is $m_c =\langle x \rangle$. The phase transition is due again to the appearance of condensation phenomena in the outcomes of the sample mean \cite{marsili2012concentration,majumdar2005nature}.

In the following sections, we recover some known results about condensation phenomena in the large-deviation regime using the so-called Density Functional Method (DFM). The DFM has been introduced in the context of Random Matrix Theory under the name of \emph{Coulomb Gas} \cite{dean2006large}, but is a very natural way to investigate the appearance of condensation phenomena in i.i.d.\ random variables. With a simple \emph{condensed ansatz}, the DFM allows us to fully characterize the condensed phase of the system and to bypass the failure of the LDT due to a vanishing rate function. In this way, we are able at once to recover the most important features of the system in the thermodynamic limit, such as the marginal distribution of the variables, the average value of the order parameter, and the phase diagram of the system. Moreover, in the democratic phase, we are able to verify the large-deviation principle and to extend the results of Sanov's theorem to random variables with fat-tailed distributions \cite{mezard2009information,cover2012elements}.

\subsection{Democratic Phase}
\label{sec:dem_phase}

Let us consider again $N$ non-negative i.i.d.\ random variables $\{x_1,x_2,\dots,x_N\}$ with sample mean $M_N$. The p.d.f.\ of $M_N$ is defined in Eq.\ (\ref{eq:partfunc}). We assume that all random variables are distributed according to a generic \emph{sub-exponential} distribution $p(x)$, i.e.\ a distribution whose tail decays slower than an exponential. Power-law distributions, such as (\ref{eq:pdf}), are included in this category. The DFM rely in the following procedure. In the thermodynamic limit $N\to\infty$, we can make a change of variables and trade the multiple integral (\ref{eq:partfunc}) over the $N$ variables $\{x_1,x_2,\dots,x_N\}$ with a functional integral over the density function:
\begin{equation}
\rho(x) = \frac{1}{N} \sum_{i=1}^N \delta(x-x_i) \ .
\label{eq:df}
\end{equation}
This leads to:
\begin{equation}
P_N(M_N=m) \propto \int\!\mathcal{D}\rho\,e^{-NE[\rho]} \; \df{\int\!\upd x\, \rho(x) -1} \df{\int\!\upd x\, x\,\rho(x) -m} \ ,
\label{eq:funcint}
\end{equation}
where the delta functions fix the normalization of $\rho(x)$ and the constraint $M_N=m$ respectively. The effective energy $E[\rho]$ reads:
\begin{equation}
E[\rho] =  \int\!\upd x\, \rho(x) \ln\frac{\rho(x)}{p(x)} \ ,
\label{eq_en_0}
\end{equation}
and is exactly the Kullback-Leibler divergence $D_\mathrm{KL}(\rho\Vert p)$ between the two distributions $\rho(x)$ and $p(x)$ \cite{mezard2009information,cover2012elements}. The functional $E[\rho]$ is obtained as the difference of two terms: the \emph{energetic term} $H[\rho]=-\int\!\upd x\,\rho(x)\ln p(x)$, which is the functional form of the Hamiltonian (\ref{eq:energy}); and the \emph{entropic term} $S[\rho]=-\int\!\upd x\,\rho(x)\ln\rho(x)$, coming from the change of variables and accounting for the exponentially large number of sequences $\{x_1,x_2,\dots,x_N\}$ corresponding to a selected density $\rho(x)$. The functional integral (\ref{eq:funcint}) can be evaluated through a saddle-point approximation. This leads to:
\begin{equation}
P_N(M_N=m) \sim e^{-NE[\rho^*]} \qquad \text{for $N\to\infty$,}
\label{eq:saddle}
\end{equation}
where the density $\rho^*(x)$ minimizes the effective energy $E[\rho]$ under the constraints expressed by the delta functions. The saddle-point density $\rho^*(x)$ can be found through the method of Lagrange multipliers, i.e.\ through the minimization of the functional:
\begin{equation}
E_\mathrm{LM}[\rho] = E[\rho] + \mu_0 \left( \int\!\upd x\, \rho(x) -1 \right) + \mu_1 \left( \int\!\upd x\, x\, \rho(x) -m \right) \ ,
\end{equation}
with respect to $\rho(x)$, $\mu_0$, and $\mu_1$. In this way, we find:
\begin{equation}
\rho^*(x) = \frac{p(x)\,e^{\mu_1 x}}{\int\!\upd x\, p(x)\, e^{\mu_1 x}}
\label{eq_rhostar}
\end{equation}
where the Lagrange multiplier $\mu_1$ is fixed by the constraint:
\begin{equation}
\frac{\int\!\upd x\, x\, p(x)\,e^{\mu_1 x}}{\int\!\upd x\, p(x)\, e^{\mu_1 x}} = m \ .
\label{eq_const_0}
\end{equation}
The solution $\rho^*(x)$ exists as long as Eq.\ (\ref{eq_const_0}) admits a solution. The asymptotic relation (\ref{eq:saddle}) proves that the system obeys the large-deviation principle (\ref{eq:LDT}), then, comparing the two equations, we are able to compute the rate function of the system, which is given by $I(m)=E[\rho^*]$ or, equivalently, $I(m)=D_\mathrm{KL}(\rho^*\Vert p)$. These findings extends the results of Sanov's theorem \cite{mezard2009information,cover2012elements} from descrete random variables to fat-tailed distributions, as long as the system is in the democratic phase (see next section).

The saddle-point density $\rho^*(x)$ is not just a mathematical tool but has a specific physical meaning, and this can be shown through the following steps. Let us denote by $\langle\;\cdot\;\rangle_m$ the average of a generic random variable according to the constrained measure (\ref{eq:joint}) and let us define the \emph{marginal distribution} of the variables at fixed mean $m$:
\begin{equation}
\big\langle\rho(x)\big\rangle_m = \left\langle \frac{1}{N} \sum_{i=1}^N \delta(x-x_i) \right\rangle_m \ .
\label{eq:marginal}
\end{equation}
With some simple algebra, starting from the definitions, we can express $\langle\rho(x)\rangle_m$ in term of the p.d.f.\ $P_N(M_N=m)$:
\begin{equation}
\big\langle\rho(x)\big\rangle_m = p(x)\cdot\frac{N}{N-1}\cdot\frac{P_{N-1}\big(M_{N-1}=\tfrac{Nm-x}{N-1}\big)}{P_N\big(M_N=m\big)} \ ,
\end{equation}
then, in the limit $N\to\infty$, we can invoke the large-deviation principle (\ref{eq:LDT}) and we obtain:
\begin{equation}
\big\langle\rho(x)\big\rangle_m \sim p(x)\, e^{-N[I(m-x/N)-I(m)]} \ .
\label{eq:temp}
\end{equation}
Using the relation $I(m)=E[\rho^*]$ with the saddle-point solutions (\ref{eq_rhostar}) and (\ref{eq_const_0}), we find that the exponent in (\ref{eq:temp}) converges to a finite value. The result is:
\begin{equation}
\big\langle\rho(x)\big\rangle_m \sim p(x)\, e^{\mu_1 x} \ ,
\end{equation}
which is exactly the non-normalized expression of the density (\ref{eq_rhostar}). Therefore, we have proved the following fundamental relation:
\begin{equation}
\lim_{N\to\infty}\big\langle\rho(x)\big\rangle_m = \rho^*(x) \ ,
\label{eq:marginal_limit}
\end{equation}
which means that the saddle point density $\rho^*(x)$ is the asymptotic form of the marginal distribution of the variables in the thermodynamic limit.

Let us make few comments about the previous results. The asymptotic relation (\ref{eq:marginal_limit}), together with Eqs.\ (\ref{eq_rhostar}) and (\ref{eq_const_0}), clearly shows the democratic behaviour of the system. It means that, in the large-deviations regime, all variables tend to be distributed according to the tilted distribution $\rho^*(x)$ instead of the original distribution $p(x)$. The tilted distribution is characterized by a shifted average value which is exactly equal to $m$, so all variables equally contribute to the large-deviations of $M_N$. As a final result, we can use the knowledge about the asymptotic density $\rho^*(x)$ to evaluate the typical behavior of the order parameter $Y_k$ at fixed $m$. It is simple to prove that:
\begin{equation}
\langle Y_k \rangle_m = \frac{1}{N^{k-1} m^k} \int\upd x\,x^k \big\langle\rho(x)\big\rangle_m \ .
\label{eq:average}
\end{equation}
Since $\langle\rho(x)\rangle_m\to\rho^*(x)$ for $N\to\infty$, the integral in (\ref{eq:average}) is always finite and so, for large $N$, the expected value $\langle Y_k\rangle_m$ vanishes. Once again, it is clear that the above results describe a system in the democratic phase, and cannot explain the behaviour of the variables in the presence of condensation phenomena.

\subsection{Condensed Phase}
\label{sec:cond_phase}

The results presented in the previous section are valid as long as Eq.\ (\ref{eq_const_0}) admits a solution. If $p(x)$ is sub-exponential and $\langle x\rangle$ is finite, then the solution exists only for $m\leq\langle x\rangle$, whereas, for $m>\langle x\rangle$, the whole procedure fails. If we use expressions (\ref{eq:funcint}) and (\ref{eq_en_0}) to define the p.d.f.\ $P_N(M_N = m)$, we are implicitly assuming that the large deviations of $M_N$ are obtained through a democratic outcome, since all variables $\{x_1,x_2,\dots,x_N\}$ have a unique scaling behaviour determined by the asymptotic density $\rho^*(x)$. The above computations, then, rely on a \emph{democratic ansatz}. In order to explore the unaccessible region $m>\langle x \rangle$, we should violate this ansatz and turn to a \emph{condensed ansatz}, imposing that variables could have different scaling behaviours. We thus go back to Eq. (\ref{eq:partfunc}) and perform a different change of variables, exploiting the incomplete density function:
\begin{equation}
\rho_c(x) = \frac{1}{N-1} \sum_{i=1}^{N-1} \delta(x-x_i) \ .
\label{eq:df_BIS}
\end{equation}
The new density describes all variable but one, namely $x_N$, which can have a different scaling behaviour with respect to the other variables. Since all variables are interchangeable, the specific choice of $x_N$ does not affect the results, and we denote the selected variable simply as $x_c$.
The new ansatz accounts for the spontaneous symmetry break $\mathcal{S}_{N}\mapsto\mathcal{S}_{N-1} \times 1$, where $\mathcal{S}_N$ is the permutation group of $N$ elements. Indeed, in the democratic phase the system is invariant under a generic permutation of the random variables, whereas in the condensed phase the appearance of a condensed variable breaks down this symmetry. Expressions (\ref{eq:funcint}) and (\ref{eq_en_0}), written in term of the density $\rho_c(x)$, now read:
\begin{equation}
\begin{split}
& P_N(M_N=m) \propto \int\!\mathcal{D}\rho_c\,\upd x_c\,e^{-NE[\rho_c,x_c]} \,\times \\
& \qquad \times \df{\int\!\upd x\, \rho_c(x) -1} \df{\frac{N-1}{N} \int\!\upd x\, x\,\rho(x) +\frac{x_c}{N} -m} \ ,\label{eq:funcint_MOD}
\end{split}
\end{equation}
and:
\begin{equation}
E[\rho_c,x_c] = \frac{N-1}{N}\int\!\upd x\, \rho_c(x)\ln\frac{\rho_c(x)}{p(x)} - \frac{1}{N}\ln p(x_c) \ .
\label{eq_en_1}
\end{equation}
In order to apply the saddle-point approximation to the functional integral in the limit $N\to\infty$, we must rescale all variables such that the leading terms in the integral are of the same order in $N$. The scaling laws of the variables are driven by the constraints in the delta functions. The only non-trivial choice, which involves both variables $\rho_c(x)$ and $x_c$ in the constraints' satisfaction, is $x_c=O(N)$. For this reason, we perform the substitution $x_c = Nt$ and neglect all sub-leading terms for $N\to\infty$. Since $p(x)$ is a sub-exponential distribution, the last term in Eq.\ (\ref{eq_en_1}) vanishes and we get:
\begin{equation}
\begin{split}
& P_N(M_N=m) \propto \int\!\mathcal{D}\rho_c\,\upd t\,e^{-N\widehat{E}[\rho_c,t]} \,\times \\
& \qquad \times \df{\int\!\upd x\, \rho_c(x) -1} \df{\int\!\upd x\, x\,\rho(x) +t -m} \ ,
\end{split}
\end{equation}
where:
\begin{equation}
\widehat{E}[\rho_c,t] = \int\!\upd x\, \rho_c(x)\ln\frac{\rho_c(x)}{p(x)} \ .
\label{eq_en_1bis}
\end{equation}
Finally, minimizing the functional:
\begin{equation}
\widehat{E}_\mathrm{LM}[\rho_c,t] = \widehat{E}[\rho_c,t] + \mu_0 \left( \int\!\upd x\, \rho(x) -1 \right) + \mu_1 \left( \int\!\upd x\, x\, \rho(x) +t -m \right) \ ,
\end{equation}
with respect to $\rho_c(x)$, $t$, $\mu_0$, and $\mu_1$, we find the saddle-point solutions $\rho_c^*(x)= p(x)$ and $t^*=m-\langle x \rangle$. Therefore, in the limit $N\to\infty$, the $N-1$ non-condensed random variables are distributed according to the original p.d.f.\ $p(x)$. Roughly speaking, they ignore the constraint $M_N=m$ and behave as i.i.d.\ random variables. At variance, the condensed variable $x_c^*=Nt^*$ scales as:
\begin{equation}
x_c^* \simeq N(m-\langle x \rangle)
\label{eq:condensate}
\end{equation} 
and carries a finite fraction of $M_N$. Notice that, at the leading order in $N$, the saddle-point energy $E[\rho_c^*,x_c^*]$ vanishes and leads to a vanishing rate function $I(m)$ for all $m>\langle x \rangle$. It means that the system does not obey the large-deviation principle (\ref{eq:LDT}) or, in a statistical-mechanics interpretation, that the free-energy of the system is a sub-extensive quantity. This result is in full agreement with the standard results of the LDT.

At this stage, we are able to evaluate the behavior of the order parameter $\langle Y_k\rangle_m$ in the condensed phase by means of Eq. (\ref{eq:average}). By comparing the two densities (\ref{eq:df}) and (\ref{eq:df_BIS}), we can write the asymptotic expression of $\langle\rho(x)\rangle_m$ in the thermodynamic limit as $\rho_c^*(x)+N^{-1}\delta(x-x_c)$. This leads to the following result:
\begin{equation}
\langle Y_k \rangle_m \simeq \left( 1- \frac{\langle x\rangle}{m} \right)^k \ ,
\label{eq:OP}
\end{equation}
which is, of course, a clear sign of condensation. Notice that such result depends only on the scaling law (\ref{eq:condensate}) and is independent on the behaviour of non-condensed variables (this is true as long as finite-$N$ fluctuations are neglected). The behavior of the average IPR $\langle Y_2 \rangle_m$ as a function of $m$ is shown in Fig.\ \ref{fig_MC} and agrees with numerical simulation (see Appendix).

\begin{figure}
\centering
\psfrag{X}[c][c]{{$m/m_c$}}
\psfrag{Y}[r][c]{{$\langle Y_2\rangle_m\!\!\!\!$}}
\includegraphics[width=0.6\columnwidth]{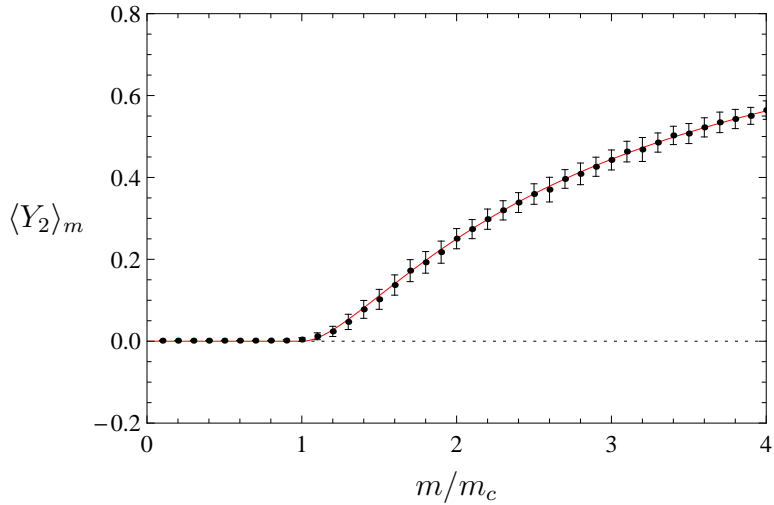}
\caption{\onehalfspacing Average value of the IPR at fixed mean for sub-exponential distributions with finite $\langle x\rangle$. The red line depicts the asymptotic formula (\ref{eq:OP}) for $N\to\infty$. The errorbars are mumerical estimations of $\langle Y_2\rangle_m$ obtained by means of Monte-Carlo simulations (see Appendix). The simulated system is composed of $N=1000$ random variables distributed according to a shifted Pareto distribution $p(x)=\alpha (x+1)^{-(\alpha+1)}$ with $x\geq0$ and $\alpha=3$.}
\label{fig_MC}
\end{figure}

\begin{figure}
\centering
\psfrag{X}[c][c]{$\alpha$}
\psfrag{Y}[r][c]{$m$}
\includegraphics[width=0.6\columnwidth]{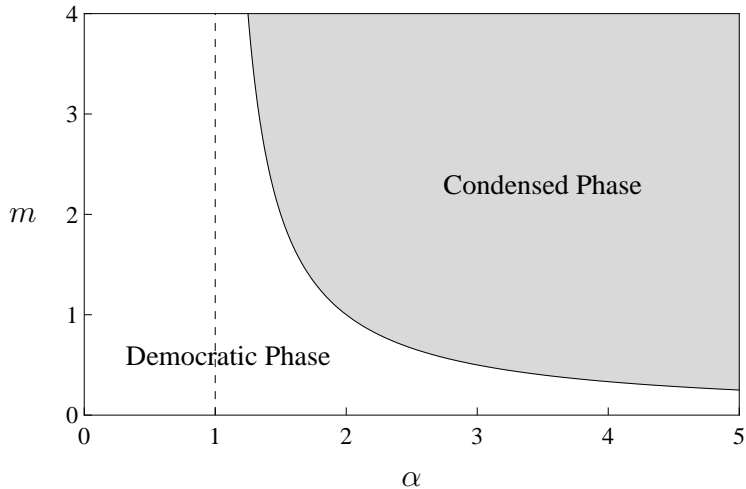}
\caption{\onehalfspacing Phase diagram of large deviations of i.i.d.\ random variables with a power-law distribution at fixed mean $m$ and tail-index $\alpha$. No condensation can be observed for $\alpha<1$. The plot has been obtained with a shifted Pareto distribution $p(x)=\alpha (x+1)^{-(\alpha+1)}$ with $x\geq0$.}
\label{fig_phdiag2}
\end{figure}

In conclusion, the DFM allows to identify the critical point $m_c=\langle x \rangle$ separating a \emph{democratic phase} ($m<m_c$) from a \emph{condensed phase} ($m>m_c$). In the thermodynamic limit, the expected value $\langle Y_k \rangle_m$ behaves as a good order parameter: it vanishes in the democratic phase and increases in the condensed phase, according to Eq. (\ref{eq:OP}). Condensation phenomena occur only for sub-exponential distributions and only if the expected value $\langle x \rangle$ is finite. These results allows to draw the phase diagram of the system. The case of power-law distributions (\ref{eq:pdf}) is depicted in Fig.\ \ref{fig_phdiag2}: it is in full agreement with both the standard LDT results \cite{marsili2012concentration} and the gran-canonical analysis of the mass-transport model performed in \cite{majumdar2005nature}. In this case we observe also an interesting phenomenon, namely, the reversion of the condensation criteria. In the large-deviation regime, indeed, the condensed phase appears only for $\alpha>1$ whereas, in the typical-fluctuations regime, it occurs for $\alpha<1$ (see Sec.\ \ref{sec:setting}). This can be explained as follows. In the typical-fluctuations regimes (i.e.\ in absence of the constraint $M_N=m$) the position of the system in the $\alpha$ -- $m$ plane would be exactly on the phase boundary between the democratic and the condensed phases, because $M_N$ would converge to the expected value $\langle x\rangle$. In the large-deviation regime, the constraint $M_N=m$ allows the system to leave the boundary line and to move in either of the two phases, but only for $\alpha>1$. For $\alpha<1$, instead, the typical value of $M_N$ diverges, so the constraint $M_N=m$ forces the system to be always below the critical point and set it in the democratic phase. Therefore, the reversion of the condensation criteria depends on the anomalous scaling law $S_N\sim N^{1/\alpha}$ for $\alpha<1$, which causes the typical value of $M_N$ to diverge. Such reversion mechanism is very general: roughly speaking, every time we add a new constraint to the system (such as $M_N=m$), we implicitly modify its scaling-laws, destroying the previous condensed phases and generating new ones. We shall onbserve again this phenomenon in Sec.\ \ref{sec:general}, where a constraint on $Y_k$ is discussed.

\section{Realized Volatility of Stock Prices}
\label{sec:prices}

Sub-exponential distributions recur very often in natural and social sciences, and our previous analytical results can be useful to describe and to understand a large variety of phenomena. In this section, we analyze the appearance of condensation phenomena in financial time-series of stock-prices. This is a very interesting scenario: condensation phenomena in finance are quite common and play a fundamental role in the evaluation of financial risk. They take the name of \emph{jumps}, \emph{crashes}, or \emph{flash-crashes}, according to their magnitude, time-scale, systemic diffusion, etc.\ In the worst cases, thay can have strong effects on worldwide economy.

\begin{figure}
\centering
\psfrag{X}[c][c]{{$M_N\,/\,\overline{M_N}$}}
\psfrag{Y}[r][c]{{$Y_2$}}
\includegraphics[width=\textwidth]{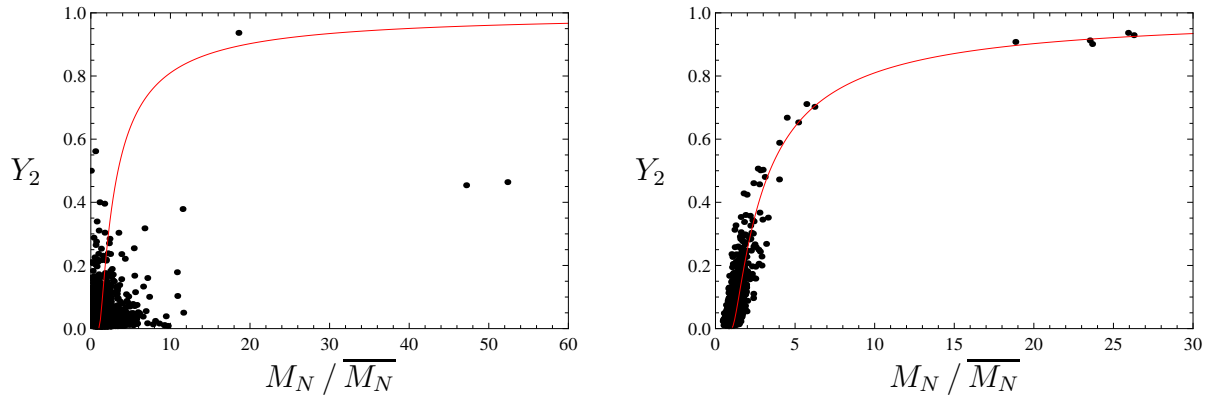}
\caption{\onehalfspacing Scatter plots of the IPR versus the realized volatility of squared price returns for the FTSE MIB 40 in the period 2012-04-01/2013-08-31. Left: real time-series. Right: reshuffled time-series. Each point represents a measurement on one stock and one trading day, for a total of 14.200 points per plot. The realized volatility $M_N$ has been rescaled by its average value $\overline{M_N}$ on each stock. The IPR has been evaluated by dividing each one-day return into $N=450$ one-minute returns. The continuous lines show the expected behaviour of the IPR in the limit $N\to\infty$ (see Eq.\ (\ref{eq:OP})).}
\label{fig_prices}
\end{figure}

In the following, we consider the 40 most traded stocks of the Italian Market (FTSE MIB 40) from April 2012 to August 2013. Our dataset is composed of the \emph{best ask} and \emph{best bid} prices, i.e.\ the lowest/highest price at wich a specific stock can be bought/sold, updated with the time-resolution of 1 second.
We perform the following analysis. We divide each trading day into time intervals of 1 minute. In order to remove the anomalies in price fluctuations at the opening and closure of the market, we exclude the first and last 30 minutes of trades from each day, leaving 450 minutes of trades per day. Then, for each time interval $[t_{i-1},t_i]$, we measure the \emph{squared logarithmic price-returns}:
\begin{equation}
x_i = \left[\log\frac{P(t_i)}{P(t_{i-1})}\right]^2 \ ,
\end{equation}
where $P(t)$ is the \emph{mid-price} of a stock at time $t$ (the mid-price is defined as the arithmetic mean of the best ask and bid prices). The squared returns $x_i$ can be considered as non-negative and fat-tailed distributed random variables, since linear price-returns are usually distributed according to power-law distibutions with a tail index in the range of 3~--~4 \cite{bouchaud2003theory}. For each stock and each day, we measure the mean $M_N$ and the $k$-th moment of the weight $Y_k$, namely:
\begin{equation}
M_N = \frac{1}{N}\sum_{i=1}^N x_i \ , \qquad\text{and}\qquad Y_k = \frac{1}{M_N^k}\sum_{i=1}^N x_i^k \ ,
\end{equation}
where $N$ is the number of time-intervals in a single day ($N=450$). In financial literature, the observable $M_N$ is called \emph{realized volatility} and quantifies the typical size of price flucutations \cite{bouchaud2003theory}. The observable $Y_k$, at variance, can be used to identify condensation phenomena in financial time-series. The occurrence of $Y_k\approx 1$ on a specific trading day denotes the presence of untipically large price fluctuations, which are responsible for the final outcome of $M_N$. We repeat several measure of $M_N$ and $Y_k$ for each stock and each trading day in our dataset (40 stocks $\times$ 355 days $=$ 14.200 observations) and we draw a scatter plot of the outcomes. Each occurence of $M_N$ has been rescaled with respect to $\overline{M_N}$, namely, the average value of $M_N$ for each stock. The result, for $k=2$, is shown in Fig.\ \ref{fig_prices} -- left, together with the expected behavior of $\langle Y_k\rangle$ described by Eq.\ (\ref{eq:OP}). The large number of measurements allows us to observe some rare events with large $M_N$ and even some condensed events with large $Y_k$, but there is no general agreement between empirical observations and analytical expectations.
The reason of this effect is that our analytical results concern i.i.d.\ random variables, while price returns $x_i$ are not independent at all. Indeed, even if linear price-returns are not auto-correlated in time, the auto-correlation of squared returns is very strong and exhibits long-memory effects \cite{bouchaud2003theory}. In order to deal with independent random variables we can perform, stock by stock, an overall reshuffling of the time-series of $x_i$. Such procedure destroys any auto-correlation in time but preserves the statistical properties of stock returns. The new results, after the reshuffling, are presented in Fig.\ \ref{fig_prices} -- right. This time, the agreement with Eq.\ (\ref{eq:OP}) is very good, and the presence of points with very large value of both $M_N$ and $Y_k$ implies that largest values of the realized volatility are actually condensed. Our result shows that, at least in principle, condensation phenomena can occur also in time-series of stock-prices, but their realization strongly depends on the auto-correlation of price returns.
We refer to \cite{marsili_inpr} for further investigation about this topic.

\section{Large Deviations under Additional Constraint}
\label{sec:general}

In Sec.\ \ref{sec:largedev} we analyzed the typical behavior of the order parameter $Y_k$ in the large-deviations regime of the mean $M_N$, i.e.\ under the constraint $M_N=m$. In this last section we move further and we investigate the full p.d.f.\ of $Y_k$ under the same constraint. As we can see, this problem will lead us to analyze the large-deviations regime of the system under a double constraint, namely, in the case where both $M_N$ and $Y_k$ are fixed. Let us consider again $N$ random variables $\{x_1,x_2,\dots,x_N\}$ distributed according to the constrained measure (\ref{eq:joint}). Since the sample mean is fixed by the constraint $M_N=m$, instead of the order parameter $Y_k$ we can consider the rescaled parameter $R_k= N^{k-1}m^kY_k$, which is an intensive variable and is equal to the $k$-th non-centered sample moment $\frac{1}{N}\sum_{i=1}^Nx_i^k$. We can write the constrained p.d.f.\ of $R_k$ as:
\begin{equation}
P_N(R_k=r_k|M_N=m) = \frac{P_N(R_k=r_k,M_N=m)}{P_N(M_N=m)}
\label{eq:condprob}
\end{equation}
where the joint probability $P_N(R_k=r_k,M_N=m)$ reads:
\begin{equation}
\begin{split}
& P_N(R_k=r_k,M_N=m) = \int\left(\prod_{i=1}^N \upd x_i\,p(x_i)\right) \,\times \\
& \qquad \times \df{\frac{1}{N}\sum_{i=1}^N x_i - m} \df{\frac{1}{N}\sum_{i=1}^N x_i^k - r_k} \ .
\label{eq:jointprob}
\end{split}
\end{equation}
The denominator $P_N(M_N=m)$ in Eq.\ (\ref{eq:condprob}) has been already studied in the previous section and plays the role of a normalization constant, so we can focus on the numerator $P_N(R_k=r_k,M_N=m)$. As for $P_N(M_N=m)$, the integral (\ref{eq:jointprob}) can be analyzed with the DFM, following the steps of Sec.\ \ref{sec:largedev}. The democratic ansatz yields the functional integral:
\begin{equation}
\begin{split}
& P_N(R_k=r_k,M_N=m) \propto \int\!\mathcal{D}\rho\,e^{-NE[\rho]} \; \df{\int\!\upd x\, \rho(x) -1} \times \\
& \qquad \times \df{\int\!\upd x\, x\,\rho(x) -m} \df{\int\!\upd x\,x^k \rho(x)-r_k} \ ,
\label{eq:funcint_GC1}
\end{split}
\end{equation}
where $E[\rho]$ is given by Eq.\ (\ref{eq_en_0}). This integral is the same of Eq.\ (\ref{eq:funcint}) with an additional constraint on the moment $R_k$. The constrained minimization of the functional $E[\rho]$ leads to the saddle-point solution:
\begin{equation}
\rho^*(x) =\frac{p(x)\,e^{\mu_1 x+\mu_k x^k}}{\int\!\upd x\,p(x)\,e^{\mu_1 x+\mu_k x^k}} \ .
\end{equation}
The Lagrange multipliers $\mu_1$ and $\mu_k$ are functions of $m$ and $r_k$ and are implicitly defined by imposing the constraints:
\begin{equation}
\frac{\int\!\upd x\,x\,p(x)\,e^{\mu_1 x+\mu_k x^k}}{\int\!\upd x\,p(x)\,e^{\mu_1 x+\mu_k x^k}} =m \ ,
\label{eq_const_1}
\end{equation}
\begin{equation}
\frac{\int\!\upd x\,x^k\,p(x)\,e^{\mu_1 x+\mu_k x^k}}{\int\!\upd x\,p(x)\,e^{\mu_1 x+\mu_k x^k}} =r_k \ .
\label{eq_const_2}
\end{equation}
As in the previous case, the democratic ansatz holds as long as Eqs.\ (\ref{eq_const_1}) and (\ref{eq_const_2}) have a solution in terms of $\mu_1$ and $\mu_k$. If the values $m$ and $r_k$ do not admit a solution for Eqs.\ (\ref{eq_const_1}) and (\ref{eq_const_2}), then we must turn from the democratic to the condensed ansatz. In this case, we find:
\begin{equation}
\begin{split}
& P_N(R_k=r_k,M_N=m) \propto \int\!\mathcal{D}\rho_c\,\upd x_c\,e^{-NE[\rho_c,x_c]} \; \df{\int\!\upd x\, \rho_c(x) -1} \times \\
& \quad \times \df{\frac{N-1}{N} \int\!\upd x\, x\,\rho_c(x) +\frac{x_c}{N} -m} \df{\frac{N-1}{N}\int\!\upd x\,x^k \rho_c(x) +\frac{x_c^k}{N} -r_k} \ ,
\label{eq:funcint_GC2}
\end{split}
\end{equation}
with the effective energy (\ref{eq_en_1}). Once again, we must rescale the variables in Eq.\ (\ref{eq:funcint_GC2}) in order to have the same scaling in $N$ for all the leading terms. This time, assuming that $k>1$, the only non-trivial choice is given by $x_c=O(N^{1/k})$. Thus, we perform the substitutions $x_c=N^{1/k}t$ and, at the leading order, we get:
\begin{equation}
\begin{split}
& P_N(R_k=r_k,M_N=m) \propto \int\!\mathcal{D}\rho_c\,\upd t\,e^{-N\widehat{E}[\rho_c,t]} \; \df{\int\!\upd x\, \rho_c(x) -1} \times \\
& \qquad \times \df{\int\!\upd x\, x\,\rho_c(x) -m} \df{\int\!\upd x\, x^k\,\rho_c(x) +t^k -r_k} \ ,
\label{eq_en_5}
\end{split}
\end{equation}
with the effective energy (\ref{eq_en_1bis}).
The saddle-point solutions $\rho_c^*(x)$ and $t^*$ of this integral are:
\begin{equation}
\rho_c^*(x) = \frac{p(x)\,e^{\mu_1 x}}{\int\!\upd x\, p(x)\, e^{\mu_1 x}} \ ,
\label{EQ_SPSOL_rhostar}
\end{equation}
\begin{equation}
t^* = \left[r_k-\frac{\int\!\upd x\, x^k\, p(x)\, e^{\mu_1 x}}{\int\!\upd x\, p(x)\, e^{\mu_1 x}}\right]^\frac{1}{k} \ ,
\label{EQ_SPSOL_tstar}
\end{equation}
where $\mu_1$ depends on $m$ and is implicitly defined by the constraint:
\begin{equation}
\frac{\int\!\upd x\, x^k\, p(x)\, e^{\mu_1 x}}{\int\!\upd x\, p(x)\, e^{\mu_1 x}} = m \ .
\label{EQ_SPSOL_mu1}
\end{equation}
Comparing these results with the ones of the previous section (see Eqs.\ (\ref{eq_rhostar}), (\ref{eq_const_0}), and (\ref{eq:condensate})), we find that the density $\rho_c^*(x)$ under the double constraint $M_N=m$ and $R_k=r_k$ (in the condensed phase) has the same expression of the density $\rho^*(x)$ under the single constraint $M_N=m$ (in the democratic phase). At variance, the condensed variable $x_c^*=N^{1/k}t^*$ scales as:
\begin{equation}
x_c^* \simeq \big[N (r_k-\langle x^k \rangle^*)\big]^\frac{1}{k} \ ,
\end{equation}
where $\langle x^k \rangle^*$ denotes the $k$-th moment of the distribution $\rho_c^*(x)$, i.e.:
\begin{equation}
\langle x^k \rangle^* = \frac{\int\!\upd x\, x^k\, p(x)\, e^{\mu_1 x}}{\int\!\upd x\, p(x)\, e^{\mu_1 x}} \ .
\end{equation}
Therefore, the condensate obeys a different scaling law ($x_c^*\sim N^{1/k}$ instead of $x_c^*\sim N$) and is driven by the constraint on the moment rather then by the constraint on the mean.

\begin{figure}
\centering
\psfrag{X}[c][c]{$m$}
\psfrag{Y}[r][c]{$r_2$}
\includegraphics[width=\textwidth]{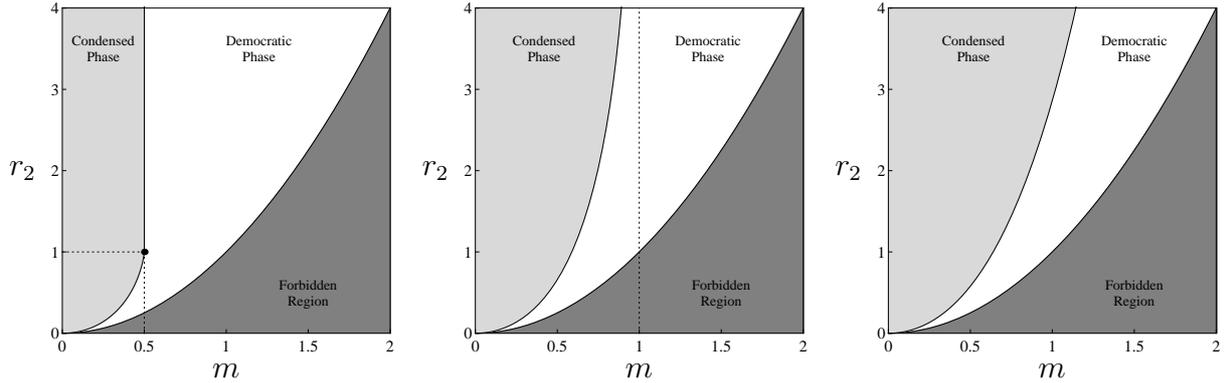}
\caption{\onehalfspacing Phase diagrams of large deviations of i.i.d.\ random variables with a sub-exponential distribution at fixed mean $m$ and 2nd moment $r_2$. Left: p.d.f.\ with finite mean and variance. Center: p.d.f.\ with finite mean and infinite variance. Right: p.d.f.\ with infinite mean and variance. The dashed lines denote the critical values $m=\langle x\rangle$ and $r_2=\langle x^2\rangle$, if any. The dark region is forbidden by the Jensen inequality: $r_2 \geq m^2$. In the case of power-law distributions (\ref{eq:pdf}), the three cases correspond to $\alpha>2$, $1<\alpha\leq2$, and $0<\alpha\leq1$, respectively. The plots have been obtained with a shifted Pareto distribution $p(x)=\alpha (x+1)^{-(\alpha+1)}$ with $x\geq0$ and $\alpha=1,2,3$.}
\label{fig_phdiag3}
\end{figure}

At this stage, we can use Eqs.\ (\ref{eq_const_1}) and (\ref{eq_const_2})  to study the phase diagram of the system in the space of the control parameters $m$ and $r_k$. The two equations should be inverted in order to write $\mu_1$ and $\mu_k$ as functions of $m$ and $r_k$. For $k>1$, the phase boundary between the democratic and the condensed phases is defined by the constraint $\mu_k(m,r_k)=0$. Indeed, for $\mu_k< 0$ the integrals in (\ref{eq_const_1}) and (\ref{eq_const_2}) diverge and the democratic ansatz does not hold anymore. Let us start by considering the case of a fat-tailed p.d.f.\ $p(x)$ with finite expectation values $\langle x \rangle$ and $\langle x^k \rangle$. For $\mu_1=0$ and $\mu_k=0$ we find the critical point $(m,r_k) = (\langle x \rangle, \langle x^k \rangle)$, which lies on the phase boundary. For $\mu_1>0$ and $\mu_k=0$ we find a regular curve which lies in the region of the space with $m < \langle x \rangle$. For $\mu_1<0$ we cannot set $\mu_k=0$ directly, but we can set $\mu_k<0$ and than take the limit $\mu_k \rightarrow 0$. We studied the last curve numerically and the results show a straight line going from $(\langle x \rangle, \langle x^k \rangle)$ to $(\langle x \rangle,\infty)$. The obtained phase diagram, for $k=2$, is represented in Fig.\ \ref{fig_phdiag3} -- left. In the case of broader distributions, such that $\langle x \rangle$ is finite but $\langle x^k \rangle$ is not, the point $(m,r_k)$ moves to $(\langle x \rangle,\infty)$, but the condensed phase remains confined in the region $m<\langle x \rangle$ (see Fig.\ \ref{fig_phdiag3} -- center). Finally, if both $\langle x \rangle$ and $\langle x^k \rangle$ diverge, the condensed phase spreads through the whole parameter space and can be observed for any value of $m$ (see Fig.\ \ref{fig_phdiag3} -- right). As noticed in Sec.\ \ref{sec:cond_phase}, we observe again the reversion of condensation criteria, from $m>\langle x\rangle$ in the case of the unique constraint $M_N=m$, to $m<\langle x\rangle$ in the case of the additional constraint $R_k=r_k$ (compare, for instance, the phase diagrams in Figs.\ \ref{fig_phdiag2} and \ref{fig_phdiag3}). When $R_k$ is not constrained, the system moves exactly on the phase boundary between the democratic and the condensed phase in the $m$ -- $r_k$ plane. The reversion of condensation criteria is due to the divergence of the typical value of $R_k$ for $m>\langle x\rangle$.

The results obtained in this section agree with the ones presented in \cite{szavits-nossan2014constraint}. We recall that these results are valid for any sub-exponential distribution and, specifically, for any power-law distribution (\ref{eq:pdf}). In this case, all our results are independent on the tail index $\alpha$, since the scaling laws of the condensate and the other non-condensed variables are driven by the constraints on $M_N$ and $R_N$, rather than by the tail of $p(x)$. The tail of the distribution enters the computations in just one (fundamental) step: it determines if the expected values $\langle x \rangle$ and $\langle x^k \rangle$ are finite, and so determines if condensation phenomena occur or not. For this reasons, our results are very general and hold in a veriety of situations, namely, for any system described by a large number of independent random variables with fat-tailed distribution. Notice, however, that our analysis is peformed in the thermodynamic limit, and neglects all fluctuations due to the finite-size $N$ of the system.

\section{Conclusions}
\label{sec:conclusions}

In this work we have investigated the underlying mechanism that is at the base of condensation phenomena in fat-tailed distibutions. Specifically, we investigated the phase transition from a democratic to a condensed phase in a generic system composed of $N$ i.i.d.\ random variables with fat-tailed distribution. The condensed phase is generated by a spontaneous symmetry-breaking mechanism and is due to the anomalous scaling-law of a single variable, namely, the condensate \cite{marsili2012concentration}. In the thermodynamic limit $N\to\infty$, the condensate carries a finite fraction of the sum $S_N$ and causes the system's rate function to vanish, yielding the failure of the standard Large Deviation Theory (in a statistical mechanics approach, this means that the thermodynamic potential of the system is non-extensive).

In this work, we have reported a thorough characterization of the phase transition in terms of an order parameter: the $k$-th moment of the weight $Y_K$ (or, specifically, the Inverse Participation Ratio $Y_2$). Such observable is non-vanishing only in the condensed phase and is non-analytical at some critical point. The study of the order parameter reveals the presence of condensed phases in different regimes, namely, in the typical-fluctuations regime, in the large-deviations regime at fixed sum $S_N$, and in the large-deviations regime where both $S_N$ and $Y_k$ are fixed. We noticed how the addition of new constraints causes a reversion of the condensation criteria, destroying previous condensed phases and generating new ones.

The characterization of the phase transition has been achieved my means of the Density Functional Method, borrowed from the field of Random Matrix Theory \cite{dean2006large}. With few simple steps, this method allows to recover the most important features of the system in the thermodynamic limit, such as the rate function, the marginal distribution, the phase diagram, and the anomalous scaling-laws of the variables in the condensed phase. The application of the Density Functional Method to the condensed phase requires a specific \emph{condensed ansatz} which explicitly accounts for the spontaneous symmetry-breaking mechanisms. The procedure allows also to extend the results of the Sanov's theorem from descrete random variables to fat-tailed distribution, as long as the system is in the democratic phase \cite{mezard2009information}.

Finally, we compared our analytical results with some numerical studies. We performed a Monte-Carlo simulation in the large-deviations regime of the sum of fat-tailed random variables by using a micro-canonical algorhitm. Our results prove the presence of a phase transition and confirm the analytical expectations on the behavoiur of the order parameter. More interestingly, we observed the occurrence of condensation phenomena in the realized volatility of stock-prices by the analysis of financial time-series from the Italian Market (FTSE MIB 40 -- from Apr.\ 2012 to Aug.\ 2013). We observed that the statistical distibution of  price returns can lead in principle to condensation phenomena, but their auto-correlation in time stongly affects their generation. A datailed study of this phenomenon is reported in \cite{marsili_inpr}.

Phase transitions due to condensation phenomena play an important role in different models. The obtained results present strong connections with the low temperature phase of disordered systems. In particular, the entropy-vanishing phase transition in the Random Energy Model is strictly related to condensation phenomena of fat-tailed random variables, as put forward in several works \cite{derrida1994non, mezard1984replica, bouchaud1997universality}. The phase transition in the large-deviations regime of $S_N$ has a direct interpretation in terms of a mass-transport model \cite{majumdar2005nature}. At variance, the phase transition in the large-deviations regime of both $S_N$ and $Y_k$ is reminiscent to the one taking place in bipartite quantum systems and concerning the distribution of the Renyi entropies \cite{nadal2010phase}.

A further discussion about condensation phenomena can be found in \cite{marsili_inpr}. Here the authors extend the analysis of this topic also to the case of non-independent random variables, such as the eigenvalues of random matrices, and discuss some applications to the financial world, such as the price-jumps in financial time series of stock prices, and the \emph{market mode} in financial correlation matrices.

\section*{Acknowledgements}

Financial support from F.S.E.\ within the framework of the S.H.A.R.M.\ P.O.R.\ 2007/2013 project and from LIST S.p.A.\ is acknowledged. Data providing from LIST S.p.A. is also acknowledged. The authors want to thank L.\ Caniparoli, G.\ Livan, and M.\ Peressi for fruitful discussions, E.\ Dameri and E.\ Melchioni for fostering the present collaboration, and D.\ Davio for the continuous encouragement. We thank S.N.\ Majumdar for the fruitful interaction.

\section*{Appendix: Monte-Carlo Simulations}

Here we present some results about condensation phenomena in fat-tailed distributions obtained through Monte-Carlo simultations. Our aim is to verify the behaviour of the order parameter $\langle Y_k\rangle_m$ expressed by Eq.\ (\ref{eq:OP}) by means of numerical simulations. The Monte-Carlo technique allows to directly investigate the large-deviations regime defined by the constraint $M_N=m$.
We adopted a \emph{micro-canonical} Metropolis algorithm, based on the following steps:
\begin{enumerate}
\item Set the initial values of the variables to a state with $M_N=m$.
\item Propose a move which leaves the total value of $M_N$ unchanged. This can be easily achieved through the substitutions $x_A\mapsto x_A+\varepsilon$ and $x_B\mapsto x_B-\varepsilon$, where $x_A$ and $x_B$ are randomly chosen variables and $\varepsilon$ is a random step.
\item Accept or reject the move according to the probability distribution $p(x)$, namely, with probability $p(x_A+\varepsilon)/p(x_A)$ times $p(x_B-\varepsilon)/p(x_B)$.
\item Repeat steps 2 and 3 until the system reaches the equilibrium.
\end{enumerate}
The dynamics of the simulated system can be very slow when passing from the democratic phase to the condensed phase, or vice versa. When the critical point is reached, the condensed variable must shift from $O(1)$ to $O(N)$ and so, if $N$ is large, the system can get stuck in a non-equilibrium state for a long time, altering the results of the simulation. The above algorithm reaches the best performance by imposing that either $x_A$ or $x_B$ is always the largest among all variables. In this way, the condensed variable is allowed to have very large fluctuations, and the system can shift between the two phases very quickly. The results obtained with this method are presented in Fig.\ \ref{fig_MC} and show a perfect agreement with Eq.\ (\ref{eq:OP}).

\bibliographystyle{unsrt}
\bibliography{bibliography}

\begin{thebibliography}{10}

\bibitem{majumdar2005nature}
S.~N. Majumdar, M.~R. Evans, and R.~K.~P. Zia.
\newblock Nature of the condensate in mass transport models.
\newblock {\em Physical Review Letters}, 94:180601, 2005.

\bibitem{marsili2012concentration}
M.~Marsili.
\newblock On the concentration of large deviations for fat tailed
  distributions.
\newblock {\em arXiv:1201.2817v1}, 2012.

\bibitem{touchette2009large}
H.~Touchette.
\newblock The large deviation approach to statistical mechanics.
\newblock {\em Physics Reports}, 478:1, 2009.

\bibitem{szavits-nossan2014constraint}
J.~Szavits-Nossan, M.~R. Evans, and S.~N. Majumdar.
\newblock Constraint-driven condensation in large fluctuations of linear
  statistics.
\newblock {\em Physical Review Letters}, 112:020602, 2014.

\bibitem{zaliapin2005approximating}
I.~V. Zaliapin, Y.~Y. Kagan, and F.~P. Schoenberg.
\newblock Approximating the distribution of pareto sums.
\newblock {\em Pure and Applied geophysics}, 162:1187, 2005.

\bibitem{saichev2006universal}
A.~Saichev and D.~Sornette.
\newblock Universal distribution of interearthquake times explained.
\newblock {\em Physical Review Letters}, 97:078501, 2006.

\bibitem{malamud1998forest}
B.~D. Malamud, G.~Morein, and D.~L. Turcotte.
\newblock Forest fires: an example of self-organized critical behavior.
\newblock {\em Science}, 281:1840, 1998.

\bibitem{peters2002rain}
O.~Peters and K.~Christensen.
\newblock Rain: Relaxations in the sky.
\newblock {\em Physical Review E}, 66:036120, 2002.

\bibitem{zipf1949human}
G.~K. Zipf.
\newblock {\em Human Behavior and the Principle of Least Effort}.
\newblock Addison-Wesley, 1949.

\bibitem{pareto1964cours}
V.~Pareto.
\newblock {\em Cours d'{\'E}conomie Politique}.
\newblock Droz, 1964.

\bibitem{mantegna1995scaling}
R.~N. Mantegna and H.~E. Stanley.
\newblock Scaling behaviour in the dynamics of an economic index.
\newblock {\em Nature}, 376:46, 1995.

\bibitem{embrechts1997modelling}
P.~Embrechts, C.~Kl{\"u}ppelberg, and T.~Mikosch.
\newblock {\em Modelling Extremal Events: For Insurance and Finance}.
\newblock Springer, 1997.

\bibitem{pitaevskii2003bose}
L.~P. Pitaevskii and S.~Stringari.
\newblock {\em Bose-Einstein Condensation}.
\newblock Clarendon Press, 2003.

\bibitem{mezard2009information}
M.~M{\'e}zard and A.~Montanari.
\newblock {\em Information, Physics, and Computation}.
\newblock Oxford University Press, 2009.

\bibitem{bouchaud2003theory}
J.-P. Bouchaud and M.~Potters.
\newblock {\em Theory of Financial Risk and Derivative Pricing: From
  Statistical Physics to Risk Management}.
\newblock Cambridge University Press, 2003.

\bibitem{nadal2010phase}
C.~Nadal, S.~N. Majumdar, and M.~Vergassola.
\newblock Phase transitions in the distribution of bipartite entanglement of a
  random pure state.
\newblock {\em Physical Review Letters}, 104:110501, 2010.

\bibitem{bianconi2001bose}
G.~Bianconi and A.~L. Barab{\'a}si.
\newblock Bose-einstein condensation in complex networks.
\newblock {\em Physical Review Letters}, 86:5632, 2001.

\bibitem{marsili_inpr}
M.~Filiasi, G.~Livan, M.~Marsili, M.~Peressi, E.~Vesselli, and E.~Zarinelli.
\newblock On the concentration of large deviations for fat-tailed
  distributions, with application to financial data.
\newblock {\em Available at SSRN 1985596}, 2014.

\bibitem{gnedenko1968limit}
B.~V. Gnedenko and A.~N. Kolmogorov.
\newblock {\em Limit Distributions for Sums of Independent Random Variables}.
\newblock Addison-Wesley, 1968.

\bibitem{bouchaud1990anomalous}
J.-P. Bouchaud and A.~Georges.
\newblock Anomalous diffusion in disordered media: statistical mechanisms,
  models and physical applications.
\newblock {\em Physics Reports}, 195:127, 1990.

\bibitem{derrida1994non}
B.~Derrida.
\newblock Non-self-averaging effects in sums of random variables, spin glasses,
  rrandom maps and random walks.
\newblock In {\em On Three Levels: Micro, Meso and Macroscopic Approaches in
  Physics}, pages 125--137. M. Fannes et al. (eds). Plenum Press, 1994.

\bibitem{dean2006large}
D.~S. Dean and S.~N. Majumdar.
\newblock Large deviations of extreme eigenvalues of random matrices.
\newblock {\em Physical Review Letters}, 97:160201, 2006.

\bibitem{cover2012elements}
T.~M. Cover and J.~A. Thomas.
\newblock {\em Elements of Information Theory}.
\newblock Wiley, 2012.

\bibitem{mezard1984replica}
M.~M{\'e}zard, G.~Parisi, N.~Sourlas, G.~Toulouse, and M.~Virasoro.
\newblock Replica symmetry breaking and the nature of the spin glass phase.
\newblock {\em Journal de Physique}, 45:843, 1984.

\bibitem{bouchaud1997universality}
J.-P. Bouchaud and M.~M{\'e}zard.
\newblock Universality classes for extreme-value statistics.
\newblock {\em Journal of Physics A: Mathematical and General}, 30:7997, 1997.

\end{thebibliography}

\end{document}